\documentclass[%
 reprint,
 amsmath,amssymb,
 prl
]{revtex4-2}

\usepackage{dcolumn}
\usepackage{amsmath,amsfonts,amssymb,bm,dsfont,lipsum,mathtools}
\usepackage{booktabs,enumerate,graphicx,scrextend,xcolor,comment}

\newcommand{\aff}{}

\DeclarePairedDelimiter{\bpar}{\Big (}{\Big )}

\newcommand{\grad}{\nabla}
\renewcommand{\div}{\nabla\cdot}

\renewcommand{\u}{\mathbf{u}}

\newcommand{\x}{\mathbf{x}}
\newcommand{\z}{\mathbf{z}}

\newcommand{\Ra}{\text{Ra}}
\renewcommand{\Re}{\text{Re}}
\renewcommand{\Pr}{\text{Pr}}
\newcommand{\St}{\text{St}}

\newcommand{\C}{^\circ\text{C}}

\graphicspath{{figures/}}

\begin{document}

\preprint{APS/123-QED}

\title{Anomalous convective flows carve pinnacles and scallops in melting ice}

\author{Scott Weady$^\aff{1}$, Joshua Tong$^\aff{1,2}$, Alexandra Zidovska$^\aff{2}$ \& Leif Ristroph$^\aff{1}$}
\email{ristroph@cims.nyu.edu}

\affiliation{$^\aff{1}$Applied Math Lab, Courant Institute, New York University, New York NY 10012 USA\\
$^\aff{2}$Department of Physics, New York University, New York NY 10003 USA}

\date{\today}
\begin{abstract}
We report on the shape dynamics of ice suspended in cold fresh water and subject to the natural convective flows generated during melting. Experiments reveal shape motifs for increasing far-field temperature: Sharp pinnacles directed downward at low temperatures, scalloped waves for intermediate temperatures between 5 and $7^\circ$C, and upward pointing pinnacles at higher temperatures. Phase-field simulations reproduce these morphologies, which are closely linked to the anomalous density-temperature profile of liquid water. Boundary layer flows yield pinnacles that sharpen with accelerating growth of tip curvature while scallops emerge from a Kelvin-Helmholtz-like instability caused by counterflowing currents that roll up to form vortex arrays. By linking the molecular-scale effects underlying water's density anomaly to the macro-scale flows that imprint the surface, these results show that the morphology of melted ice is a sensitive indicator of ambient temperature.

\end{abstract}

\maketitle


The shape of a landform or landscape holds clues to its history and the environmental conditions under which it developed. However, interpreting geological morphologies is challenging due to the complex multi-scale and interactive processes involved, such as erosion and deposition, dissolution and solidification, and melting and freezing \cite{Huang:2015,Mullins:1963,Wettlaufer:1997,Ristroph:2012,Hewett:2017,Nakouzi:2015}. The latter yield examples across scales, including rippled icicles, pinnacle shaped icebergs, textured ice caves, and larger icescapes \cite{Curl:1966,Kobayashi:1980,Neufeld:2010,Camporeale:2012,Romanov:2012,Filhol:2015}. Understanding how to interpret such forms and the physical mechanisms behind them is all the more important due to the increasing melt rate of the Earth's ice reserves \cite{chen2006satellite,chen2009accelerated}.

Melting is an example of a Stefan problem, which classically seeks to determine interface motion induced by a phase transition \cite{Rubenstein:1971}. Here the solid-liquid interface recedes due to temperature gradients normal to the surface, and the energy released during phase change in turn modifies the temperature field in the fluid. In many situations, these temperature changes cause density variations that drive gravitational convective flows, which also feed back on the interface motion \cite{Ristroph:2018,Pegler:2021}. This convective Stefan problem has recently been studied in the related context of solids dissolving into liquids, where the effects of flows due to solutal convection can be seen in fine-scale surface features and overall forms \cite{Davies-Wykes:2018,Huang:2020,Pegler:2020}. 

\begin{figure}[b!]
\centering \vspace{-0.3cm}
\includegraphics[scale=0.60]{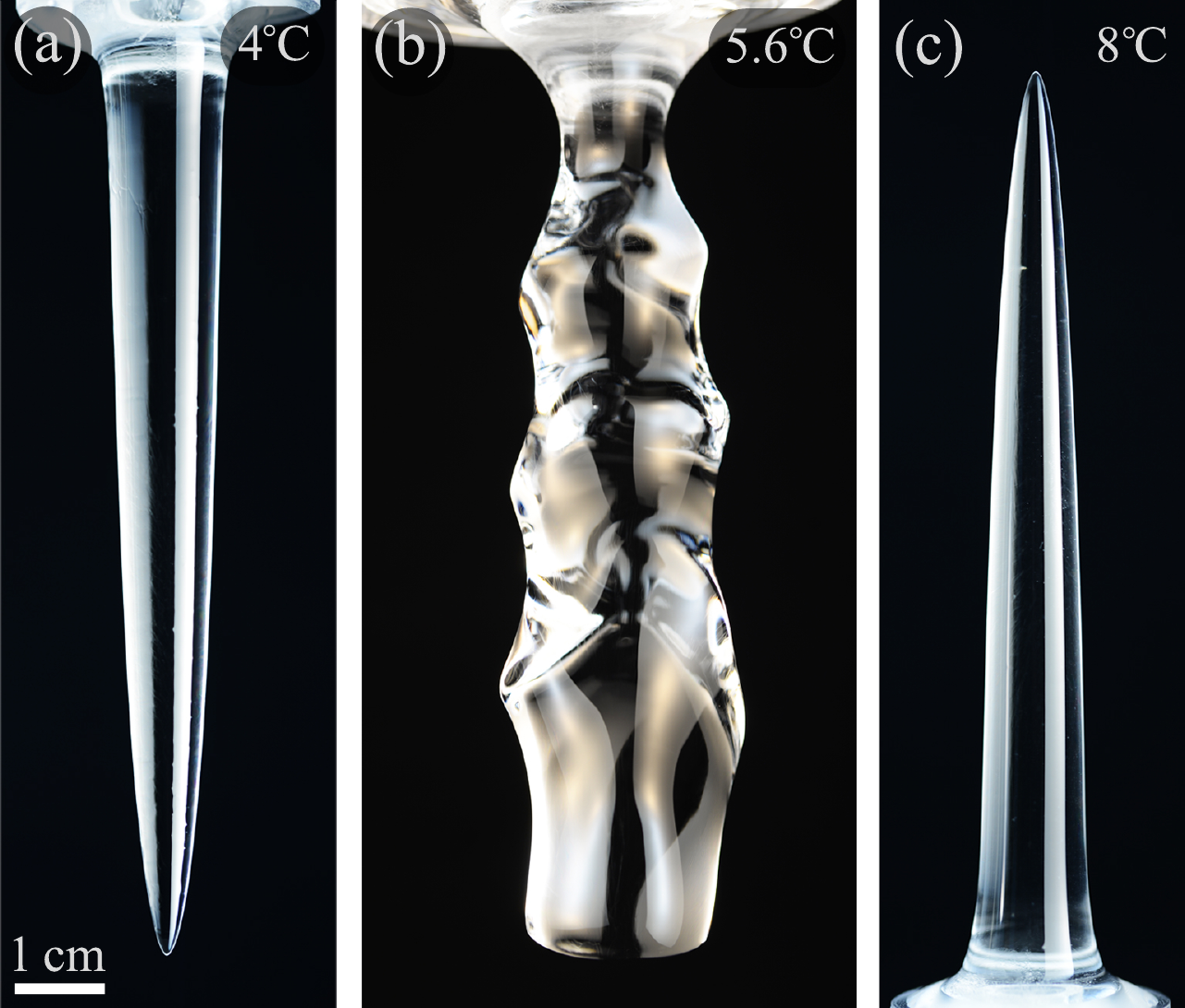}\vspace{-0.3cm}
\caption{Representative morphologies formed by melting ice in laboratory experiments. (a) For sufficiently cold ambient temperatures $T_\infty \lesssim 5\C$, the ice tapers from below to form an inverted pinnacle. (b) For intermediate temperatures $5\C\lesssim T_\infty\lesssim7\C$, scalloped patterns form on the surface. (c) Upright pinnacles form for warmer temperatures $T_\infty \gtrsim 7\C$. Photographs capture the late-stage ice removed from water and under diffuse lighting.}\label{fig:experiments}
\end{figure}

Convective flows are well studied in heat transfer problems involving fixed boundaries \cite{Tritton:1977,Schlichting:2017}, but the effects of shape-flow coupling are less understood. The problem is uniquely complex for melting ice due to the unusual effect of temperature on liquid water's density, which displays a maximum at about $4\C$. This so-called density anomaly, while ultimately a molecular-scale effect that leads to relative density differences on the order of $0.01\%$ \cite{Nilsson:2015}, nonetheless strongly affects convective flows, heat transfer characteristics, and hydrodynamic instabilities across a wide range of scales \cite{Veronis:1962,Higgins:1983,Topp:2018,Bendell:1976,Saitoh:1976,Gebhart:1982,White:1986,Siegert:2021,Qi:2021}.

\begin{figure*}[t!]
\centering
\includegraphics[scale=1.00]{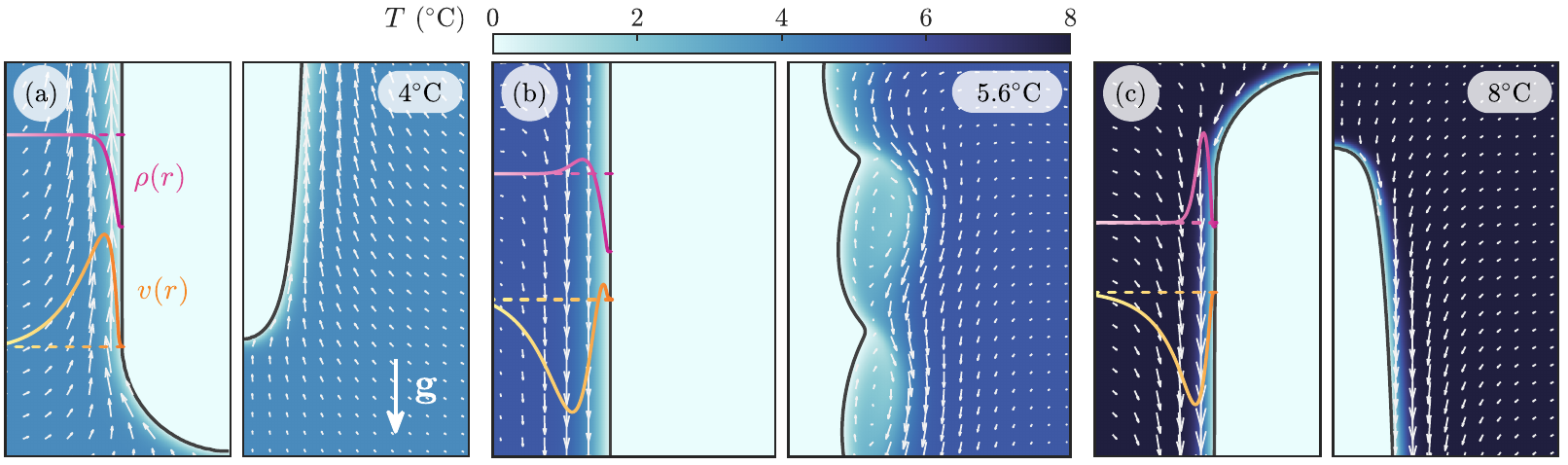} \vspace{-0.4cm}
\caption{Flow velocity (arrows) and temperature (color) fields from phase-field simulations for (a) $T_\infty = 4\C$, (b) $5.6\C$, and (c) $8\C$ at early (left) and late (right) times. Curves show the flow speed (yellow) and density (pink) profiles at early times. Melting in cold water is associated with upward flow and downwards tapering of the ice as in (a), whereas warmer temperatures involve downwards flow and tapering at the top as in (c). Intermediate temperatures yield shear flows involving rising fluid near the surface and sinking outer flows, driving an instability that later patterns the surface.}\label{fig:flows}\vspace{-0.5cm}
\end{figure*}


Here we show that the density anomaly and consequent flows are imprinted onto the shape of melting ice. We consider the highly simplified context of ice submerged within fresh water while subject to the convective flows generated during melting. Reporting first on experiments, we manufacture clear ice using a directional freezing method \cite{carte1961air,maeno1967air}, and immerse it in water of fixed far-field temperature $T_\infty \in [2,10]\C$.  We focus on cylindrical initial forms that are sized, supported, and oriented vertically to allow for observation of the long-time shape dynamics. The ice is rigidly supported underwater on a plastic base that is located either at the top or bottom of the ice depending on the ambient water temperature. Using a large tank in a cold room facility, the far-field water temperature is controlled and systematically varied to assess its impact on shape development, as captured by time-lapse photography. Prior to melting in water, nearly uniform internal temperature of $0\C$ is achieved by leaving the ice in room temperature air for at least 30 minutes, which is long compared to the timescale of thermal diffusion. Experimental details are available as Supplemental Material.
 
We first present some motivating observations from experiments, which reveal three distinct morphologies that arise for specific intervals of the far-field temperature. Representative photographs are shown in Fig. \ref{fig:experiments}. For sufficiently low temperatures $T_\infty \lesssim 5\C$, the ice becomes tapered at its lower end to form an inverted pinnacle with its apex pointing downward, as shown in panel (a). For such conditions, the base is at the top of the ice, which avoids interference with the upward boundary layer flows to be discussed below. At higher temperatures $T_\infty \gtrsim 7\C$, a similarly shaped but upright pinnacle forms, shown in (c), where the base is at the bottom of the ice. For intermediate temperatures $5\C\lesssim T_\infty \lesssim 7\C$, intricate wavy and scalloped features pattern the ice surface, as shown in (b).


The sensitive dependence of shape on temperature evokes water's anomalous density-temperature profile, whose peak at $T_* \approx 4\C$ suggests distinct scenarios categorized by the far-field temperature $T_\infty$. For sufficiently warm temperatures, the cold liquid near the surface is uniformly denser than that in the far-field and is expected to sink. For cold temperatures, however, the density anomaly upends intuition: Cold liquid near the surface is less dense and will rise. Intermediate temperatures are more subtle, since the coldest fluid near the surface is less dense than that in the far field, while fluid slightly further away must be at or near $T_*$ and is thus more dense. The resulting flows are not easily inferred.

\begin{figure*}[t!]
\centering
\includegraphics[scale=1.00]{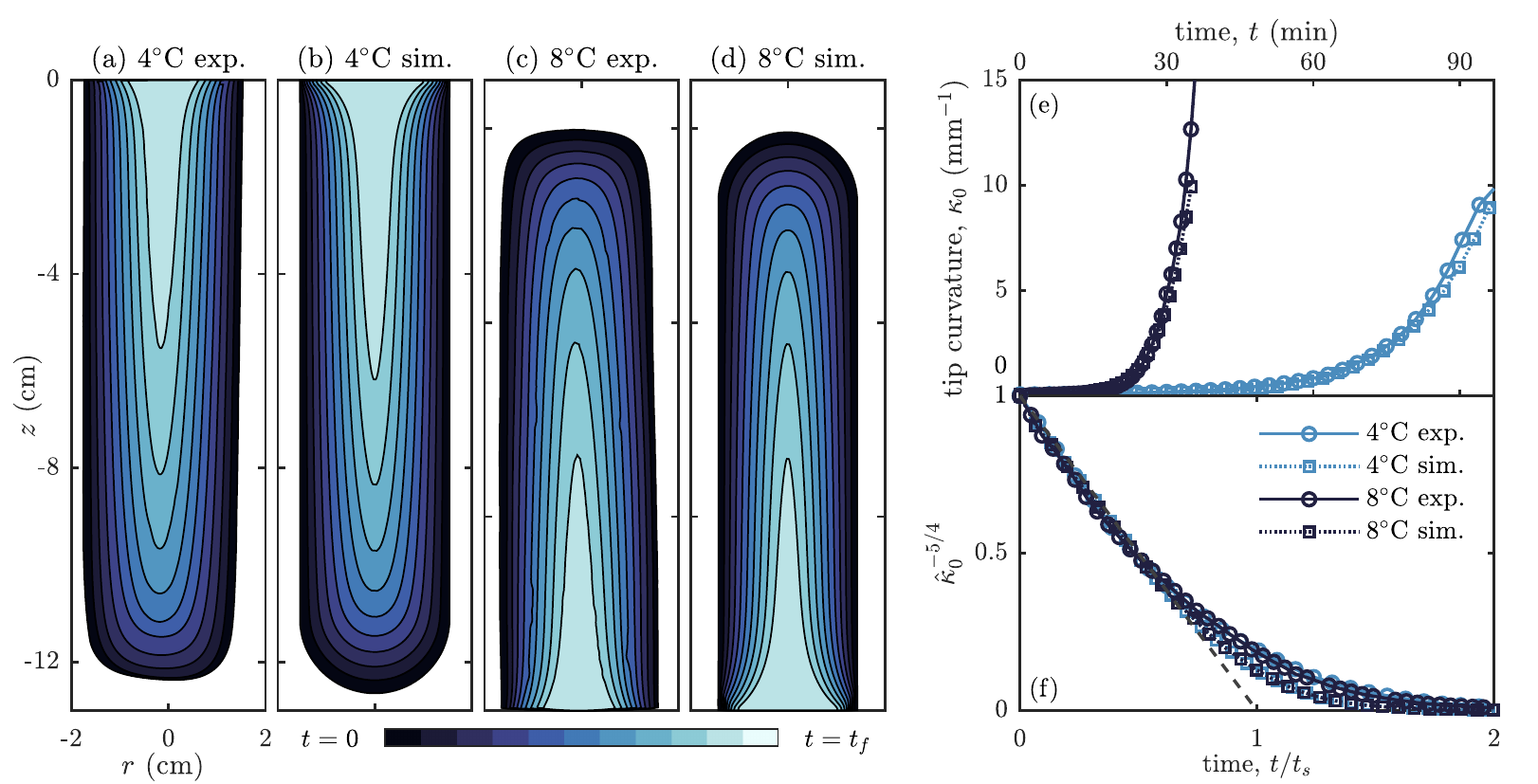}\vspace{-0.4cm}
\caption{Dynamics of pinnacle formation. Panels (a)-(d) show interfaces for $T_\infty = 4\C$ and $T_\infty = 8\C$ extracted over time (dark to light blue) in experiments and simulations. The corresponding tip curvatures are plotted in (e), which exhibit rapid sharpening to micro-scales. At early times, all data follow the scaling law $\kappa_0(t) = \kappa_0(0)(1 - t/t_s)^{-4/5}$, as shown in (f) by the linear behavior $1-t/t_s$ of the transformed quantity $[\hat\kappa_0(t/t_s)]^{-5/4} = [\kappa_0(t/t_s)/\kappa_0(0)]^{-5/4}$.}\label{fig:spikes}
\end{figure*}


To more clearly interpret the observed morphologies, we formulate and implement simulations of the shape dynamics coupled to the natural convective flows. Here we use the phase-field model \cite{Wang:1993,Beckermann:1999}, which has proven successful for moving boundary problems with natural convection \cite{Favier:2019,Couston:2021,Hester:2021}. In this model, material is implicitly represented by a continuous phase parameter $\phi(\x,t)$, which takes values $\phi = 0$ in the solid phase and $\phi = 1$ in the liquid, with the interface defined as the level set $\phi = 1/2$. This phase parameter is then used to describe energy contributions from phase change and to approximate the no-slip boundary condition on the ice, while admitting numerical discretization on a Cartesian grid. Introducing the quadratic equation of state $\rho(T) = \rho_*[1 - \beta(T - T_*)^2]$ as a basic model for the density anomaly \cite{Veronis:1962,Topp:2018}, the fluid motion is described by the Navier-Stokes equation in the Boussinesq approximation
\begin{gather}
\frac{D\u}{Dt} = \Pr\bpar{-\grad p + \Delta\u + \Ra\theta^2 \hat\z} - \eta(1 - \phi)^2\u,\label{eq:du/dt}\\
\div\u = 0,\label{eq:divu}
\end{gather}
where $\u(\x,t)$ is the velocity, $p(\x,t)$ is the pressure, and ${\theta(\x,t) = (T(\x,t) - T_*)/(T_\infty - T_0)}$ is the dimensionless temperature. Parameters include the Rayleigh number ${\Ra = g\beta(T_\infty - T_0)^2H^3/\nu\kappa_T}$, which compares buoyant and viscous forces, and the Prandtl number ${\Pr = \nu/\kappa_T}$, which assesses viscous and thermal diffusivities. Here $g$ is acceleration due to gravity, $H$ is the initial height of the ice, $\nu$ is the fluid viscosity, and $\kappa_T$ is the thermal diffusivity. The last term in (\ref{eq:du/dt}) is a Brinkman penalization force that models the ice as a porous medium, in which the velocity vanishes for large resistivity $\eta\gg 1$ \cite{Angot:1999}. 

Following thermodynamic derivations \cite{Favier:2019}, the temperature and phase fields satisfy the evolution equations
\begin{gather}
\frac{D\theta}{Dt} = \Delta\theta - \frac{1}{\St}\frac{df}{d\phi}\frac{\partial\phi}{\partial t},\label{eq:dT/dt}\\
\frac{\partial\phi}{\partial t} = m\Delta\phi + \frac{m(\theta - \theta_0)}{\delta^2}\frac{df}{d\phi} - \frac{m}{4\delta^2}\frac{dg}{d\phi}.\label{eq:dphi/dt}
\end{gather}
The last term in Eq. (\ref{eq:dT/dt}) captures energy contributions from phase change, the magnitude of which is controlled by the Stefan number $\St = c_p(T_\infty - T_0)/\mathcal L$, with $c_p$ the heat capacity and $\mathcal L$ the latent heat of fusion. In Eq. (\ref{eq:dphi/dt}), $m$ is a regularization parameter, $\delta$ is an effective interface thickness, $\theta_0 = (T_0 - T_*)/(T_\infty-T_0)$ is the dimensionless melting temperature, and the functions $f(\phi) = \phi^3(10 - 15\phi + 5\phi^2)$ and $g(\phi) = \phi^2(1 - \phi)^2$ are potentials that ensure no phase-change occurs away from the interface. In the limit $\delta\rightarrow0$ and $\eta\rightarrow\infty$, the system (\ref{eq:du/dt})-(\ref{eq:dphi/dt}) recovers the Navier-Stokes equations with no-slip boundary conditions on the ice and the Stefan condition for the interface velocity $V_n = \St~ \partial \theta / \partial n$, where the temperature is assumed to be $T_0$ throughout the solid \cite{Favier:2019}.


For comparison with the cylindrical geometries in experiments, we solve the system of equations (\ref{eq:du/dt})-(\ref{eq:dphi/dt}) on an axisymmetric domain sufficiently wide such that the far-field temperature remains constant to within $0.1\C$. Estimated from the experimental parameters, we take $\Pr = 12$ and $\Ra/T_\infty^2 = 2.5\times10^6~(\C)^{-2}$. The selected Stefan number $\St/T_\infty = 0.05 ~ (\C)^{-1}$ is larger than the experimental value $\St/T_\infty = 0.012 ~ (\C)^{-1}$, which reduces simulation run time while having negligible effect on the shape dynamics. Upon initialization, the temperature in the solid phase is set to $T_0$ while the temperature in the liquid is $T_\infty$. See the Supplemental Material for implementation details, including spatial and temporal discretization \cite{Brown:2001}.

Figure \ref{fig:flows} shows the interfaces at early and late times for three representative values of $T_\infty$. Also shown are the flow velocity and temperature fields as well as curves representing the density and velocity profiles along a horizontal transect at early times, the latter substantiating our inferences based on the density anomaly. For the case $T_\infty = 4\C$ of (a), an upward boundary layer flow persists for all times. This flow is fed by warmer outer fluid that is continuously entrained from the sides and bottom, offering a mechanism for the enhanced melt rate that tapers the ice from below. For $T_\infty = 8\C$, shown in (c), the situation is similar but inverted: A downward boundary layer flow tapers the top of the ice to form an upright pinnacle. For the intermediate case $T_\infty = 5.6\C$ of (b), early times are marked by a thin region of upward flow near the surface surrounded by a broader region of downward flow. This shear flow eventually destabilizes and forms recirculating vortices that entrain the warmer outer fluid and carve scallop-shaped indentations. Across all cases, stagnation points of the flow are associated with sharp features of the surface, including the apexes of the pinnacles and cusped crests of the scalloped waves.

\begin{figure*}[t!]
\centering
\includegraphics[scale=1.00]{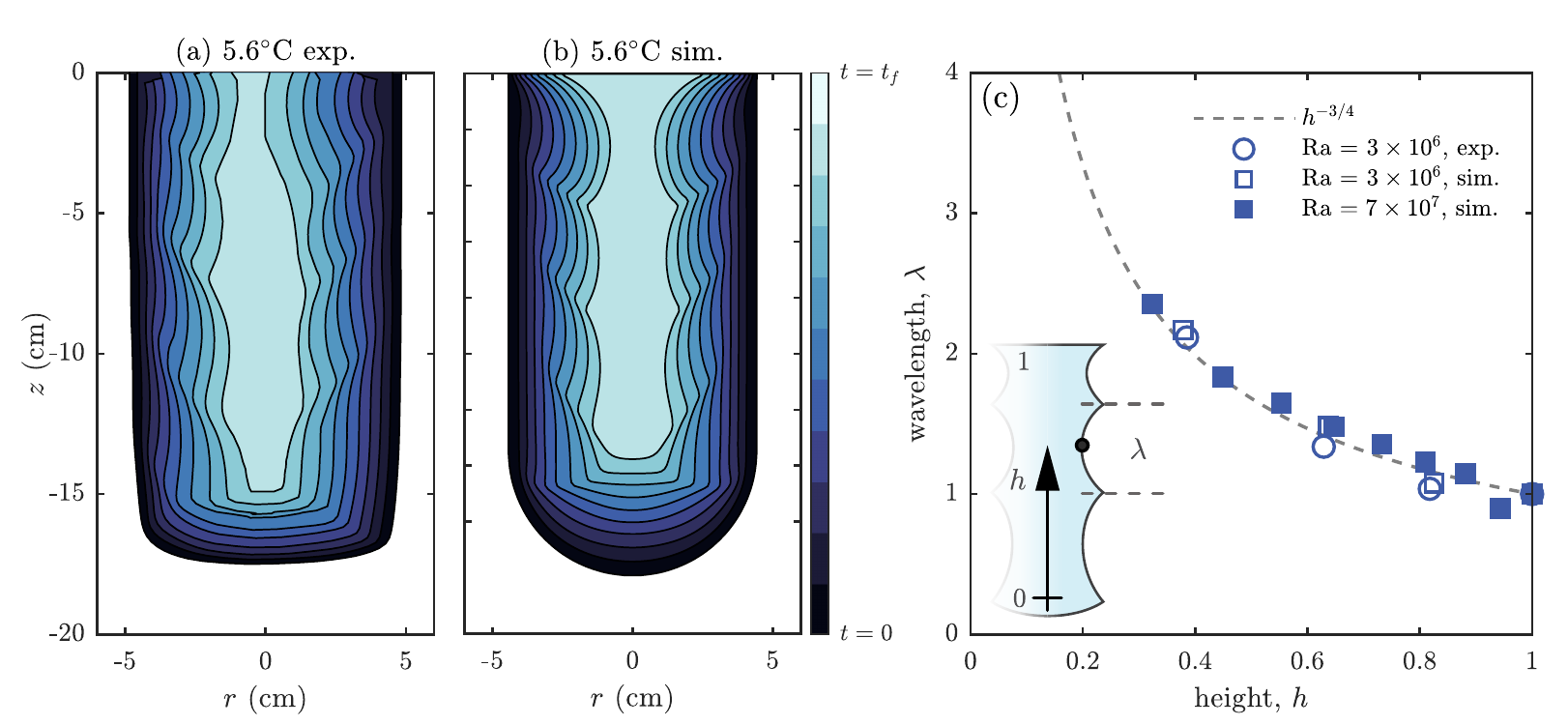}\vspace{-0.4cm}
\caption{Dynamics of scallop formation. Panels (a) and (b) show extracted interfaces over time for $T_\infty = 5.6\C$ in experiment and simulation, respectively. Panel (c) plots the cusp-to-cusp wavelength of the scallops versus their vertical location or height, confirming the scaling $\lambda \sim h^{-3/4}$ predicted by an analysis of the viscous Kelvin-Helmholtz instability. Additional simulations (filled squares) for a taller body and thus higher $\Ra$ confirm the trend over wider ranges of the variables.}\label{fig:scallops}
\end{figure*}


Further analysis of the experiments and simulations provides insight into the mathematical structure of the shape dynamics. Considering first the pinnacles observed for sufficiently low or high $T_\infty$, we show in Fig. \ref{fig:spikes}(a)-(d) comparisons of the shape progression as measured in experiments and computed in simulations for $T_\infty = 4\C$ and $8\C$. The strong agreement across all times serves as a cross-validation of the simulations and experiments and demonstrates the robustness of the pinnacle form. These pinnacles are reminiscent of those recently observed for bodies dissolving in natural convective flows \cite{Nakouzi:2015,Davies-Wykes:2018,Huang:2020,Pegler:2021}, for which a boundary layer theory analysis predicts that the pinnacle apex sharpens via a power law growth of curvature: $\kappa_0(t) = \kappa_0(0)(1 - t/t_s)^{-4/5}$ \cite{Schlichting:2017,Huang:2020}. Here $\kappa_0(0)$ is the initial tip curvature and $t_s$ is the blow-up time for the predicted singular dynamics, which were shown to accurately describe the initial stages of sharpening \cite{Huang:2020}.

To test this law, we extract the interfaces over time from experiments and simulations and evaluate the apex curvature by fitting and differentiating a fourth-order polynomial for the tip height as a function of radius. Strikingly, the tip curvature $\kappa_0(t)$, shown in Fig. \ref{fig:spikes}(e), exhibits steep and seemingly unbounded growth. The radius of curvature reaches values smaller than 100 microns, as fine as a human hair and approaching the resolutions of the experiments and simulations. In panel (f) we plot the rescaled quantity $[\hat\kappa_0(t/t_s)]^{-5/4} = [\kappa_0(t/t_s)/\kappa_0(0)]^{-5/4}$, where $t_s$ is treated as a fitting parameter based on the predicted power law. Remarkably, all data collapse to the predicted linear form $1 - t/t_s$ (dashed line) for early times, indicating a mechanism shared with dissolution for the formation of ultra-sharp structures. The curvature continues to grow at later times but falls off the singular pace, an effect also observed for dissolution pinnacles and which is the subject of recent studies \cite{Huang:2020, Huang:2021}.


Experiments and simulations are also in agreement for intermediate temperatures, yielding scallops of comparable scales as seen in Figs. \ref{fig:scallops}(a) and (b) for $T_\infty = 5.6\C$. Some disparities are expected as the simulations are axisymmetric while the patterns are three-dimensional in experiments, with the images of (a) representing cross-sectional views. Nonetheless, the wave-like structures common to both are indicative of a hydrodynamic instability. Shear flows of the form observed in Fig. \ref{fig:flows}(b) are known to undergo the Kelvin-Helmholtz instability, the classic analysis of which involves counter-flowing layers of inviscid fluid \cite{Chandrasekhar:1961,Drazin:2004}. Such flows are unstable, with the smallest wavelengths growing at the fastest rates. Viscosity, however, suppresses high frequency modes, yielding a most unstable wavelength $\lambda \sim \Re^{-1/2}$, where $\Re$ is the Reynolds number \cite{Moore:1978,Dhanak:1994}. While $\Re$ is somewhat poorly defined since the flows accelerate along the surface, scaling arguments predict $\Re \sim \Ra^{1/2}$ for $\Ra\gg 1$ \cite{Grossmann:2000}, with experimental evidence suggesting an exponent slightly below $1/2$ \cite{Qiu:2001,Grossmann:2002}. Observing that $\Ra \sim h^3$, where $h$ is the vertical distance from the bottom of the ice, we predict the wavelength decreases up the surface of the ice as $\lambda \sim h^{-3/4}$.

This scaling law can be used to test the hypothesis that the maximally unstable wavelength of the shear flow is imprinted on the ice in the form of scalloped waves. We determine the locations of the wave crests at early times and identify the wavelength $\lambda$ with differences between successive crests and the height $h$ with their midpoint. Figure \ref{fig:scallops}(c) shows results from experiments (open circles) and simulations (open squares) at $T_\infty = 5.6\C$. Here, $h$ is normalized by the total height and $\lambda$ by its value at $h=1$, and the location of $h=0$ is treated as a fitting parameter due to the ambiguous location of the lowest scallop. The uppermost scallop, whose longer wavelength seems to arise from the top boundary conditions, is excluded from this analysis. Over the few wavelengths present, the data indeed follow the $-3/4$ scaling law. Additional simulations of taller ice (filled squares), details of which are given in the Supplemental Material, yield more wavelengths and correspondingly more convincing agreement.


These findings show that the shape of ice is a sensitive indicator of the ambient temperature at which it melted. Sharply-pointed pinnacles directed downwards for $T_\infty \lesssim 5\C$ and upwards for $T_\infty \gtrsim 7\C$ are formed by persistent and unidirectional boundary layer flows that rise in the former case and sink in the latter. This latter case parallels the upright pinnacles carved by downward flows observed during dissolution \cite{Nakouzi:2015,Davies-Wykes:2018,Pegler:2020,Huang:2020,Pegler:2021}, which is expected by the analogous mathematical descriptions of thermal and solutal convection. It remains for future studies to determine if the shape dynamics are quantitatively different for melting and dissolving due to the differences in their microscale physics. The scalloped waves observed here for $5\C\lesssim T_\infty \lesssim 7\C$ have their origin in bidirectional flows due to the buoyant rise of cold water near the surface and the sinking of warmer water further outward in the boundary layer. The resulting shear flows undergo a Kelvin-Helmholtz instability and roll up into vortices that carve pits in the surface. 

Pinnacles are commonly observed on icebergs and have been qualitatively attributed to buoyancy-driven flows \cite{Romanov:2012,Moon:2018,brostrom2009iceberg}. However, they have not previously been reproduced in laboratory experiments, nor validated through fluid dynamical models or simulations. Scalloped patterns on icebergs, ice shelves, and bore holes are generally attributed to instabilities due to externally-driven flows \cite{Curl:1966,Claudin:2017,Bushuk:2019}. In contrast, the mechanism revealed here is rooted in the intrinsic flows generated by water's anomalous density characteristics, and scallops formed in this way can be distinguished by their increasing wavelength with depth. While our results pertain strictly to fresh water, the identified shape motifs may persist in saltwater solutions up to the critical concentration above which the anomaly is lost and density decreases monotonically with temperature \cite{Higgins:1983}. Future studies that vary both far-field temperature and salinity should assess how the long-time shape dynamics is impacted by the associated double-diffusive processes \cite{Higgins:1983,Josberger:1981}.

We are grateful for support from an NSF graduate fellowship to S.W., an NYU WiPhy fellowship to J.T., and NSF grants PHY-1554880 to A.Z. and CBET-1805506 and DMS-1646339 to L.R.

\bibliography{ice}

\begin{thebibliography}{54}%
\makeatletter
\providecommand \@ifxundefined [1]{%
 \@ifx{#1\undefined}
}%
\providecommand \@ifnum [1]{%
 \ifnum #1\expandafter \@firstoftwo
 \else \expandafter \@secondoftwo
 \fi
}%
\providecommand \@ifx [1]{%
 \ifx #1\expandafter \@firstoftwo
 \else \expandafter \@secondoftwo
 \fi
}%
\providecommand \natexlab [1]{#1}%
\providecommand \enquote  [1]{``#1''}%
\providecommand \bibnamefont  [1]{#1}%
\providecommand \bibfnamefont [1]{#1}%
\providecommand \citenamefont [1]{#1}%
\providecommand \href@noop [0]{\@secondoftwo}%
\providecommand \href [0]{\begingroup \@sanitize@url \@href}%
\providecommand \@href[1]{\@@startlink{#1}\@@href}%
\providecommand \@@href[1]{\endgroup#1\@@endlink}%
\providecommand \@sanitize@url [0]{\catcode `\\12\catcode `\$12\catcode
  `\&12\catcode `\#12\catcode `\^12\catcode `\_12\catcode `\%12\relax}%
\providecommand \@@startlink[1]{}%
\providecommand \@@endlink[0]{}%
\providecommand \url  [0]{\begingroup\@sanitize@url \@url }%
\providecommand \@url [1]{\endgroup\@href {#1}{\urlprefix }}%
\providecommand \urlprefix  [0]{URL }%
\providecommand \Eprint [0]{\href }%
\providecommand \doibase [0]{http://dx.doi.org/}%
\providecommand \selectlanguage [0]{\@gobble}%
\providecommand \bibinfo  [0]{\@secondoftwo}%
\providecommand \bibfield  [0]{\@secondoftwo}%
\providecommand \translation [1]{[#1]}%
\providecommand \BibitemOpen [0]{}%
\providecommand \bibitemStop [0]{}%
\providecommand \bibitemNoStop [0]{.\EOS\space}%
\providecommand \EOS [0]{\spacefactor3000\relax}%
\providecommand \BibitemShut  [1]{\csname bibitem#1\endcsname}%
\let\auto@bib@innerbib\@empty
\bibitem [{\citenamefont {Huang}\ \emph {et~al.}(2015)\citenamefont {Huang},
  \citenamefont {Moore},\ and\ \citenamefont {Ristroph}}]{Huang:2015}%
  \BibitemOpen
  \bibfield  {author} {\bibinfo {author} {\bibfnamefont {J.~M.}\ \bibnamefont
  {Huang}}, \bibinfo {author} {\bibfnamefont {M.~N.~J.}\ \bibnamefont {Moore}},
  \ and\ \bibinfo {author} {\bibfnamefont {L.}~\bibnamefont {Ristroph}},\
  }\href {\doibase 10.1017/jfm.2014.718} {\bibfield  {journal} {\bibinfo
  {journal} {Journal of Fluid Mechanics}\ }\textbf {\bibinfo {volume} {765}},\
  \bibinfo {pages} {R3} (\bibinfo {year} {2015})}\BibitemShut {NoStop}%
\bibitem [{\citenamefont {Mullins}\ and\ \citenamefont
  {Sekerka}(1963)}]{Mullins:1963}%
  \BibitemOpen
  \bibfield  {author} {\bibinfo {author} {\bibfnamefont {W.~W.}\ \bibnamefont
  {Mullins}}\ and\ \bibinfo {author} {\bibfnamefont {R.~F.}\ \bibnamefont
  {Sekerka}},\ }\href {\doibase 10.1063/1.1702607} {\bibfield  {journal}
  {\bibinfo  {journal} {Journal of Applied Physics}\ }\textbf {\bibinfo
  {volume} {34}},\ \bibinfo {pages} {323} (\bibinfo {year} {1963})}\BibitemShut
  {NoStop}%
\bibitem [{\citenamefont {Wettlaufer}\ \emph {et~al.}(1997)\citenamefont
  {Wettlaufer}, \citenamefont {Worster},\ and\ \citenamefont
  {Huppert}}]{Wettlaufer:1997}%
  \BibitemOpen
  \bibfield  {author} {\bibinfo {author} {\bibfnamefont {J.~S.}\ \bibnamefont
  {Wettlaufer}}, \bibinfo {author} {\bibfnamefont {M.~G.}\ \bibnamefont
  {Worster}}, \ and\ \bibinfo {author} {\bibfnamefont {H.~E.}\ \bibnamefont
  {Huppert}},\ }\href {\doibase 10.1017/S0022112097006022} {\bibfield
  {journal} {\bibinfo  {journal} {Journal of Fluid Mechanics}\ }\textbf
  {\bibinfo {volume} {344}},\ \bibinfo {pages} {291–316} (\bibinfo {year}
  {1997})}\BibitemShut {NoStop}%
\bibitem [{\citenamefont {Ristroph}\ \emph {et~al.}(2012)\citenamefont
  {Ristroph}, \citenamefont {Moore}, \citenamefont {Childress}, \citenamefont
  {Shelley},\ and\ \citenamefont {Zhang}}]{Ristroph:2012}%
  \BibitemOpen
  \bibfield  {author} {\bibinfo {author} {\bibfnamefont {L.}~\bibnamefont
  {Ristroph}}, \bibinfo {author} {\bibfnamefont {M.~N.~J.}\ \bibnamefont
  {Moore}}, \bibinfo {author} {\bibfnamefont {S.}~\bibnamefont {Childress}},
  \bibinfo {author} {\bibfnamefont {M.~J.}\ \bibnamefont {Shelley}}, \ and\
  \bibinfo {author} {\bibfnamefont {J.}~\bibnamefont {Zhang}},\ }\href
  {\doibase 10.1073/pnas.1212286109} {\bibfield  {journal} {\bibinfo  {journal}
  {Proceedings of the National Academy of Sciences}\ }\textbf {\bibinfo
  {volume} {109}},\ \bibinfo {pages} {19606} (\bibinfo {year}
  {2012})}\BibitemShut {NoStop}%
\bibitem [{\citenamefont {Hewett}\ and\ \citenamefont
  {Sellier}(2017)}]{Hewett:2017}%
  \BibitemOpen
  \bibfield  {author} {\bibinfo {author} {\bibfnamefont {J.~N.}\ \bibnamefont
  {Hewett}}\ and\ \bibinfo {author} {\bibfnamefont {M.}~\bibnamefont
  {Sellier}},\ }\href {\doibase
  https://doi.org/10.1016/j.jfluidstructs.2017.01.011} {\bibfield  {journal}
  {\bibinfo  {journal} {Journal of Fluids and Structures}\ }\textbf {\bibinfo
  {volume} {70}},\ \bibinfo {pages} {295} (\bibinfo {year} {2017})}\BibitemShut
  {NoStop}%
\bibitem [{\citenamefont {Nakouzi}\ \emph {et~al.}(2015)\citenamefont
  {Nakouzi}, \citenamefont {Goldstein},\ and\ \citenamefont
  {Steinbock}}]{Nakouzi:2015}%
  \BibitemOpen
  \bibfield  {author} {\bibinfo {author} {\bibfnamefont {E.}~\bibnamefont
  {Nakouzi}}, \bibinfo {author} {\bibfnamefont {R.~E.}\ \bibnamefont
  {Goldstein}}, \ and\ \bibinfo {author} {\bibfnamefont {O.}~\bibnamefont
  {Steinbock}},\ }\href {\doibase 10.1021/la503562z} {\bibfield  {journal}
  {\bibinfo  {journal} {Langmuir : the ACS journal of surfaces and colloids}\
  }\textbf {\bibinfo {volume} {31}} (\bibinfo {year} {2015}),\
  10.1021/la503562z}\BibitemShut {NoStop}%
\bibitem [{\citenamefont {Curl}(1966)}]{Curl:1966}%
  \BibitemOpen
  \bibfield  {author} {\bibinfo {author} {\bibfnamefont {R.}~\bibnamefont
  {Curl}},\ }\href@noop {} {\bibfield  {journal} {\bibinfo  {journal} {Trans.
  Cave Res. Group 7}\ ,\ \bibinfo {pages} {121}} (\bibinfo {year}
  {1966})}\BibitemShut {NoStop}%
\bibitem [{\citenamefont {Kobayashi}(1980)}]{Kobayashi:1980}%
  \BibitemOpen
  \bibfield  {author} {\bibinfo {author} {\bibfnamefont {S.}~\bibnamefont
  {Kobayashi}},\ }\href@noop {} {\bibfield  {journal} {\bibinfo  {journal}
  {Contributions from the Institute of Low Temperature Science}\ }\textbf
  {\bibinfo {volume} {A29}},\ \bibinfo {pages} {1} (\bibinfo {year}
  {1980})}\BibitemShut {NoStop}%
\bibitem [{\citenamefont {Neufeld}\ \emph {et~al.}(2010)\citenamefont
  {Neufeld}, \citenamefont {Goldstein},\ and\ \citenamefont
  {Worster}}]{Neufeld:2010}%
  \BibitemOpen
  \bibfield  {author} {\bibinfo {author} {\bibfnamefont {J.~A.}\ \bibnamefont
  {Neufeld}}, \bibinfo {author} {\bibfnamefont {R.~E.}\ \bibnamefont
  {Goldstein}}, \ and\ \bibinfo {author} {\bibfnamefont {M.~G.}\ \bibnamefont
  {Worster}},\ }\href {\doibase 10.1017/S0022112009993910} {\bibfield
  {journal} {\bibinfo  {journal} {Journal of Fluid Mechanics}\ }\textbf
  {\bibinfo {volume} {647}},\ \bibinfo {pages} {287–308} (\bibinfo {year}
  {2010})}\BibitemShut {NoStop}%
\bibitem [{\citenamefont {Camporeale}\ and\ \citenamefont
  {Ridolfi}(2012)}]{Camporeale:2012}%
  \BibitemOpen
  \bibfield  {author} {\bibinfo {author} {\bibfnamefont {C.}~\bibnamefont
  {Camporeale}}\ and\ \bibinfo {author} {\bibfnamefont {L.}~\bibnamefont
  {Ridolfi}},\ }\href {\doibase 10.1017/jfm.2011.540} {\bibfield  {journal}
  {\bibinfo  {journal} {Journal of Fluid Mechanics}\ }\textbf {\bibinfo
  {volume} {694}},\ \bibinfo {pages} {225–251} (\bibinfo {year}
  {2012})}\BibitemShut {NoStop}%
\bibitem [{\citenamefont {Romanov}\ \emph {et~al.}(2012)\citenamefont
  {Romanov}, \citenamefont {Romanova},\ and\ \citenamefont
  {Romanov}}]{Romanov:2012}%
  \BibitemOpen
  \bibfield  {author} {\bibinfo {author} {\bibfnamefont {Y.~A.}\ \bibnamefont
  {Romanov}}, \bibinfo {author} {\bibfnamefont {N.~A.}\ \bibnamefont
  {Romanova}}, \ and\ \bibinfo {author} {\bibfnamefont {P.}~\bibnamefont
  {Romanov}},\ }\href {\doibase 10.1017/S0954102011000538} {\bibfield
  {journal} {\bibinfo  {journal} {{A}ntarctic Science}\ }\textbf {\bibinfo
  {volume} {24}},\ \bibinfo {pages} {77–87} (\bibinfo {year}
  {2012})}\BibitemShut {NoStop}%
\bibitem [{\citenamefont {Filhol}\ and\ \citenamefont
  {Sturm}(2015)}]{Filhol:2015}%
  \BibitemOpen
  \bibfield  {author} {\bibinfo {author} {\bibfnamefont {S.}~\bibnamefont
  {Filhol}}\ and\ \bibinfo {author} {\bibfnamefont {M.}~\bibnamefont {Sturm}},\
  }\href {\doibase https://doi.org/10.1002/2015JF003529} {\bibfield  {journal}
  {\bibinfo  {journal} {Journal of Geophysical Research: Earth Surface}\
  }\textbf {\bibinfo {volume} {120}},\ \bibinfo {pages} {1645} (\bibinfo {year}
  {2015})}\BibitemShut {NoStop}%
\bibitem [{\citenamefont {Chen}\ \emph {et~al.}(2006)\citenamefont {Chen},
  \citenamefont {Wilson},\ and\ \citenamefont {Tapley}}]{chen2006satellite}%
  \BibitemOpen
  \bibfield  {author} {\bibinfo {author} {\bibfnamefont {J.}~\bibnamefont
  {Chen}}, \bibinfo {author} {\bibfnamefont {C.}~\bibnamefont {Wilson}}, \ and\
  \bibinfo {author} {\bibfnamefont {B.}~\bibnamefont {Tapley}},\ }\href@noop {}
  {\bibfield  {journal} {\bibinfo  {journal} {Science}\ }\textbf {\bibinfo
  {volume} {313}},\ \bibinfo {pages} {1958} (\bibinfo {year}
  {2006})}\BibitemShut {NoStop}%
\bibitem [{\citenamefont {Chen}\ \emph {et~al.}(2009)\citenamefont {Chen},
  \citenamefont {Wilson}, \citenamefont {Blankenship},\ and\ \citenamefont
  {Tapley}}]{chen2009accelerated}%
  \BibitemOpen
  \bibfield  {author} {\bibinfo {author} {\bibfnamefont {J.}~\bibnamefont
  {Chen}}, \bibinfo {author} {\bibfnamefont {C.}~\bibnamefont {Wilson}},
  \bibinfo {author} {\bibfnamefont {D.}~\bibnamefont {Blankenship}}, \ and\
  \bibinfo {author} {\bibfnamefont {B.}~\bibnamefont {Tapley}},\ }\href@noop {}
  {\bibfield  {journal} {\bibinfo  {journal} {Nature Geoscience}\ }\textbf
  {\bibinfo {volume} {2}},\ \bibinfo {pages} {859} (\bibinfo {year}
  {2009})}\BibitemShut {NoStop}%
\bibitem [{\citenamefont {Rubenstein}(1971)}]{Rubenstein:1971}%
  \BibitemOpen
  \bibfield  {author} {\bibinfo {author} {\bibfnamefont {L.~I.}\ \bibnamefont
  {Rubenstein}},\ }\href@noop {} {\emph {\bibinfo {title} {The {S}tefan
  Problem}}}\ (\bibinfo  {publisher} {American Mathematical Society},\ \bibinfo
  {year} {1971})\BibitemShut {NoStop}%
\bibitem [{\citenamefont {Ristroph}(2018)}]{Ristroph:2018}%
  \BibitemOpen
  \bibfield  {author} {\bibinfo {author} {\bibfnamefont {L.}~\bibnamefont
  {Ristroph}},\ }\href {\doibase 10.1017/jfm.2017.890} {\bibfield  {journal}
  {\bibinfo  {journal} {Journal of Fluid Mechanics}\ }\textbf {\bibinfo
  {volume} {838}},\ \bibinfo {pages} {1–4} (\bibinfo {year}
  {2018})}\BibitemShut {NoStop}%
\bibitem [{\citenamefont {Pegler}\ and\ \citenamefont
  {Davies~Wykes}(2021)}]{Pegler:2021}%
  \BibitemOpen
  \bibfield  {author} {\bibinfo {author} {\bibfnamefont {S.~S.}\ \bibnamefont
  {Pegler}}\ and\ \bibinfo {author} {\bibfnamefont {M.~S.}\ \bibnamefont
  {Davies~Wykes}},\ }\href {\doibase 10.1017/jfm.2021.86} {\bibfield  {journal}
  {\bibinfo  {journal} {Journal of Fluid Mechanics}\ }\textbf {\bibinfo
  {volume} {915}},\ \bibinfo {pages} {A86} (\bibinfo {year}
  {2021})}\BibitemShut {NoStop}%
\bibitem [{\citenamefont {Davies~Wykes}\ \emph {et~al.}(2018)\citenamefont
  {Davies~Wykes}, \citenamefont {Huang}, \citenamefont {Hajjar},\ and\
  \citenamefont {Ristroph}}]{Davies-Wykes:2018}%
  \BibitemOpen
  \bibfield  {author} {\bibinfo {author} {\bibfnamefont {M.~S.}\ \bibnamefont
  {Davies~Wykes}}, \bibinfo {author} {\bibfnamefont {J.~M.}\ \bibnamefont
  {Huang}}, \bibinfo {author} {\bibfnamefont {G.~A.}\ \bibnamefont {Hajjar}}, \
  and\ \bibinfo {author} {\bibfnamefont {L.}~\bibnamefont {Ristroph}},\ }\href
  {\doibase 10.1103/PhysRevFluids.3.043801} {\bibfield  {journal} {\bibinfo
  {journal} {Phys. Rev. Fluids}\ }\textbf {\bibinfo {volume} {3}},\ \bibinfo
  {pages} {043801} (\bibinfo {year} {2018})}\BibitemShut {NoStop}%
\bibitem [{\citenamefont {Huang}\ \emph {et~al.}(2020)\citenamefont {Huang},
  \citenamefont {Tong}, \citenamefont {Shelley},\ and\ \citenamefont
  {Ristroph}}]{Huang:2020}%
  \BibitemOpen
  \bibfield  {author} {\bibinfo {author} {\bibfnamefont {J.~M.}\ \bibnamefont
  {Huang}}, \bibinfo {author} {\bibfnamefont {J.}~\bibnamefont {Tong}},
  \bibinfo {author} {\bibfnamefont {M.}~\bibnamefont {Shelley}}, \ and\
  \bibinfo {author} {\bibfnamefont {L.}~\bibnamefont {Ristroph}},\ }\href
  {\doibase 10.1073/pnas.2001524117} {\bibfield  {journal} {\bibinfo  {journal}
  {Proceedings of the National Academy of Sciences}\ }\textbf {\bibinfo
  {volume} {117}},\ \bibinfo {pages} {23339} (\bibinfo {year}
  {2020})}\BibitemShut {NoStop}%
\bibitem [{\citenamefont {Pegler}\ and\ \citenamefont
  {Davies~Wykes}(2020)}]{Pegler:2020}%
  \BibitemOpen
  \bibfield  {author} {\bibinfo {author} {\bibfnamefont {S.~S.}\ \bibnamefont
  {Pegler}}\ and\ \bibinfo {author} {\bibfnamefont {M.~S.}\ \bibnamefont
  {Davies~Wykes}},\ }\href {\doibase 10.1017/jfm.2020.507} {\bibfield
  {journal} {\bibinfo  {journal} {Journal of Fluid Mechanics}\ }\textbf
  {\bibinfo {volume} {900}},\ \bibinfo {pages} {A35} (\bibinfo {year}
  {2020})}\BibitemShut {NoStop}%
\bibitem [{\citenamefont {Tritton}(1977)}]{Tritton:1977}%
  \BibitemOpen
  \bibfield  {author} {\bibinfo {author} {\bibfnamefont {D.~J.}\ \bibnamefont
  {Tritton}},\ }\href@noop {} {\emph {\bibinfo {title} {Physical Fluid
  Dynamics}}}\ (\bibinfo  {publisher} {Springer Netherlands},\ \bibinfo {year}
  {1977})\BibitemShut {NoStop}%
\bibitem [{\citenamefont {Schlichting}\ and\ \citenamefont
  {Gersten}(2017)}]{Schlichting:2017}%
  \BibitemOpen
  \bibfield  {author} {\bibinfo {author} {\bibfnamefont {H.}~\bibnamefont
  {Schlichting}}\ and\ \bibinfo {author} {\bibfnamefont {K.}~\bibnamefont
  {Gersten}},\ }\href@noop {} {\emph {\bibinfo {title} {Boundary layer
  theory}}}\ (\bibinfo  {publisher} {Springer, Berlin, Heidelberg},\ \bibinfo
  {year} {2017})\BibitemShut {NoStop}%
\bibitem [{\citenamefont {Nilsson}\ and\ \citenamefont
  {Pettersson}(2015)}]{Nilsson:2015}%
  \BibitemOpen
  \bibfield  {author} {\bibinfo {author} {\bibfnamefont {A.}~\bibnamefont
  {Nilsson}}\ and\ \bibinfo {author} {\bibfnamefont {L.~G.~M.}\ \bibnamefont
  {Pettersson}},\ }\href {\doibase 10.1038/ncomms9998} {\bibfield  {journal}
  {\bibinfo  {journal} {Nature Communications}\ }\textbf {\bibinfo {volume}
  {6}} (\bibinfo {year} {2015}),\ 10.1038/ncomms9998}\BibitemShut {NoStop}%
\bibitem [{\citenamefont {{Veronis}}(1963)}]{Veronis:1962}%
  \BibitemOpen
  \bibfield  {author} {\bibinfo {author} {\bibfnamefont {G.}~\bibnamefont
  {{Veronis}}},\ }\href {\doibase 10.1086/147538} {\bibfield  {journal}
  {\bibinfo  {journal} {Astrophysical Journal}\ }\textbf {\bibinfo {volume}
  {137}},\ \bibinfo {pages} {641} (\bibinfo {year} {1963})}\BibitemShut
  {NoStop}%
\bibitem [{\citenamefont {Higgins}\ and\ \citenamefont
  {Gebhart}(1983)}]{Higgins:1983}%
  \BibitemOpen
  \bibfield  {author} {\bibinfo {author} {\bibfnamefont {J.~M.}\ \bibnamefont
  {Higgins}}\ and\ \bibinfo {author} {\bibfnamefont {B.}~\bibnamefont
  {Gebhart}},\ }\href {\doibase 10.1115/1.3245660} {\bibfield  {journal}
  {\bibinfo  {journal} {Journal of Heat Transfer}\ }\textbf {\bibinfo {volume}
  {105}},\ \bibinfo {pages} {767} (\bibinfo {year} {1983})}\BibitemShut
  {NoStop}%
\bibitem [{\citenamefont {Toppaladoddi}\ and\ \citenamefont
  {Wettlaufer}(2018)}]{Topp:2018}%
  \BibitemOpen
  \bibfield  {author} {\bibinfo {author} {\bibfnamefont {S.}~\bibnamefont
  {Toppaladoddi}}\ and\ \bibinfo {author} {\bibfnamefont {J.~S.}\ \bibnamefont
  {Wettlaufer}},\ }\href {\doibase 10.1103/PhysRevFluids.3.043501} {\bibfield
  {journal} {\bibinfo  {journal} {Phys. Rev. Fluids}\ }\textbf {\bibinfo
  {volume} {3}},\ \bibinfo {pages} {043501} (\bibinfo {year}
  {2018})}\BibitemShut {NoStop}%
\bibitem [{\citenamefont {Bendell}\ and\ \citenamefont
  {Gebhart}(1976)}]{Bendell:1976}%
  \BibitemOpen
  \bibfield  {author} {\bibinfo {author} {\bibfnamefont {M.~S.}\ \bibnamefont
  {Bendell}}\ and\ \bibinfo {author} {\bibfnamefont {B.}~\bibnamefont
  {Gebhart}},\ }\href {\doibase https://doi.org/10.1016/0017-9310(76)90138-1}
  {\bibfield  {journal} {\bibinfo  {journal} {International Journal of Heat and
  Mass Transfer}\ }\textbf {\bibinfo {volume} {19}},\ \bibinfo {pages} {1081}
  (\bibinfo {year} {1976})}\BibitemShut {NoStop}%
\bibitem [{\citenamefont {Saitoh}(1976)}]{Saitoh:1976}%
  \BibitemOpen
  \bibfield  {author} {\bibinfo {author} {\bibfnamefont {T.}~\bibnamefont
  {Saitoh}},\ }\href {\doibase 10.1007/BF00385849} {\bibfield  {journal}
  {\bibinfo  {journal} {Applied Scientific Research}\ }\textbf {\bibinfo
  {volume} {32}} (\bibinfo {year} {1976}),\ 10.1007/BF00385849}\BibitemShut
  {NoStop}%
\bibitem [{\citenamefont {Gebhart}\ and\ \citenamefont
  {Wang}(1982)}]{Gebhart:1982}%
  \BibitemOpen
  \bibfield  {author} {\bibinfo {author} {\bibfnamefont {B.}~\bibnamefont
  {Gebhart}}\ and\ \bibinfo {author} {\bibfnamefont {T.}~\bibnamefont {Wang}},\
  }\href {\doibase 10.1080/00986448208910908} {\bibfield  {journal} {\bibinfo
  {journal} {Chemical Engineering Communications}\ }\textbf {\bibinfo {volume}
  {13}},\ \bibinfo {pages} {197} (\bibinfo {year} {1982})}\BibitemShut
  {NoStop}%
\bibitem [{\citenamefont {White}\ \emph {et~al.}(1986)\citenamefont {White},
  \citenamefont {Viskanta},\ and\ \citenamefont {Leidenfrost}}]{White:1986}%
  \BibitemOpen
  \bibfield  {author} {\bibinfo {author} {\bibfnamefont {D.}~\bibnamefont
  {White}}, \bibinfo {author} {\bibfnamefont {R.}~\bibnamefont {Viskanta}}, \
  and\ \bibinfo {author} {\bibfnamefont {W.}~\bibnamefont {Leidenfrost}},\
  }\href {\doibase 10.1007/BF00280268} {\bibfield  {journal} {\bibinfo
  {journal} {Experiments in Fluids}\ }\textbf {\bibinfo {volume} {4}},\
  \bibinfo {pages} {171} (\bibinfo {year} {1986})}\BibitemShut {NoStop}%
\bibitem [{\citenamefont {Couston}\ and\ \citenamefont
  {Siegert}(2021)}]{Siegert:2021}%
  \BibitemOpen
  \bibfield  {author} {\bibinfo {author} {\bibfnamefont {L.-A.}\ \bibnamefont
  {Couston}}\ and\ \bibinfo {author} {\bibfnamefont {M.}~\bibnamefont
  {Siegert}},\ }\href {\doibase 10.1126/sciadv.abc3972} {\bibfield  {journal}
  {\bibinfo  {journal} {Science Advances}\ }\textbf {\bibinfo {volume} {7}},\
  \bibinfo {pages} {eabc3972} (\bibinfo {year} {2021})}\BibitemShut {NoStop}%
\bibitem [{\citenamefont {Wang}\ \emph {et~al.}(2021)\citenamefont {Wang},
  \citenamefont {Reiter}, \citenamefont {Lohse},\ and\ \citenamefont
  {Shishkina}}]{Qi:2021}%
  \BibitemOpen
  \bibfield  {author} {\bibinfo {author} {\bibfnamefont {Q.}~\bibnamefont
  {Wang}}, \bibinfo {author} {\bibfnamefont {P.}~\bibnamefont {Reiter}},
  \bibinfo {author} {\bibfnamefont {D.}~\bibnamefont {Lohse}}, \ and\ \bibinfo
  {author} {\bibfnamefont {O.}~\bibnamefont {Shishkina}},\ }\href {\doibase
  10.1103/PhysRevFluids.6.063502} {\bibfield  {journal} {\bibinfo  {journal}
  {Phys. Rev. Fluids}\ }\textbf {\bibinfo {volume} {6}},\ \bibinfo {pages}
  {063502} (\bibinfo {year} {2021})}\BibitemShut {NoStop}%
\bibitem [{\citenamefont {Carte}(1961)}]{carte1961air}%
  \BibitemOpen
  \bibfield  {author} {\bibinfo {author} {\bibfnamefont {A.~E.}\ \bibnamefont
  {Carte}},\ }\href@noop {} {\bibfield  {journal} {\bibinfo  {journal}
  {Proceedings of the Physical Society (1958-1967)}\ }\textbf {\bibinfo
  {volume} {77}},\ \bibinfo {pages} {757} (\bibinfo {year} {1961})}\BibitemShut
  {NoStop}%
\bibitem [{\citenamefont {Maeno}(1967)}]{maeno1967air}%
  \BibitemOpen
  \bibfield  {author} {\bibinfo {author} {\bibfnamefont {N.}~\bibnamefont
  {Maeno}},\ }\href@noop {} {\bibfield  {journal} {\bibinfo  {journal} {Physics
  of Snow and Ice: proceedings}\ }\textbf {\bibinfo {volume} {1}},\ \bibinfo
  {pages} {207} (\bibinfo {year} {1967})}\BibitemShut {NoStop}%
\bibitem [{\citenamefont {Wang}\ \emph {et~al.}(1993)\citenamefont {Wang},
  \citenamefont {Sekerka}, \citenamefont {Wheeler}, \citenamefont {Murray},
  \citenamefont {Coriell}, \citenamefont {Braun},\ and\ \citenamefont
  {McFadden}}]{Wang:1993}%
  \BibitemOpen
  \bibfield  {author} {\bibinfo {author} {\bibfnamefont {S.-L.}\ \bibnamefont
  {Wang}}, \bibinfo {author} {\bibfnamefont {R.}~\bibnamefont {Sekerka}},
  \bibinfo {author} {\bibfnamefont {A.}~\bibnamefont {Wheeler}}, \bibinfo
  {author} {\bibfnamefont {B.}~\bibnamefont {Murray}}, \bibinfo {author}
  {\bibfnamefont {S.}~\bibnamefont {Coriell}}, \bibinfo {author} {\bibfnamefont
  {R.}~\bibnamefont {Braun}}, \ and\ \bibinfo {author} {\bibfnamefont
  {G.}~\bibnamefont {McFadden}},\ }\href {\doibase
  https://doi.org/10.1016/0167-2789(93)90189-8} {\bibfield  {journal} {\bibinfo
   {journal} {Physica D: Nonlinear Phenomena}\ }\textbf {\bibinfo {volume}
  {69}},\ \bibinfo {pages} {189} (\bibinfo {year} {1993})}\BibitemShut
  {NoStop}%
\bibitem [{\citenamefont {Beckermann}\ \emph {et~al.}(1999)\citenamefont
  {Beckermann}, \citenamefont {Diepers}, \citenamefont {Steinbach},
  \citenamefont {Karma},\ and\ \citenamefont {Tong}}]{Beckermann:1999}%
  \BibitemOpen
  \bibfield  {author} {\bibinfo {author} {\bibfnamefont {C.}~\bibnamefont
  {Beckermann}}, \bibinfo {author} {\bibfnamefont {H.-J.}\ \bibnamefont
  {Diepers}}, \bibinfo {author} {\bibfnamefont {I.}~\bibnamefont {Steinbach}},
  \bibinfo {author} {\bibfnamefont {A.}~\bibnamefont {Karma}}, \ and\ \bibinfo
  {author} {\bibfnamefont {X.}~\bibnamefont {Tong}},\ }\href {\doibase
  https://doi.org/10.1006/jcph.1999.6323} {\bibfield  {journal} {\bibinfo
  {journal} {Journal of Computational Physics}\ }\textbf {\bibinfo {volume}
  {154}},\ \bibinfo {pages} {468} (\bibinfo {year} {1999})}\BibitemShut
  {NoStop}%
\bibitem [{\citenamefont {Favier}\ \emph {et~al.}(2019)\citenamefont {Favier},
  \citenamefont {Purseed},\ and\ \citenamefont {Duchemin}}]{Favier:2019}%
  \BibitemOpen
  \bibfield  {author} {\bibinfo {author} {\bibfnamefont {B.}~\bibnamefont
  {Favier}}, \bibinfo {author} {\bibfnamefont {J.}~\bibnamefont {Purseed}}, \
  and\ \bibinfo {author} {\bibfnamefont {L.}~\bibnamefont {Duchemin}},\ }\href
  {\doibase 10.1017/jfm.2018.773} {\bibfield  {journal} {\bibinfo  {journal}
  {Journal of Fluid Mechanics}\ }\textbf {\bibinfo {volume} {858}},\ \bibinfo
  {pages} {437–473} (\bibinfo {year} {2019})}\BibitemShut {NoStop}%
\bibitem [{\citenamefont {Couston}\ \emph {et~al.}(2021)\citenamefont
  {Couston}, \citenamefont {Hester}, \citenamefont {Favier}, \citenamefont
  {Taylor}, \citenamefont {Holland},\ and\ \citenamefont
  {Jenkins}}]{Couston:2021}%
  \BibitemOpen
  \bibfield  {author} {\bibinfo {author} {\bibfnamefont {L.-A.}\ \bibnamefont
  {Couston}}, \bibinfo {author} {\bibfnamefont {E.}~\bibnamefont {Hester}},
  \bibinfo {author} {\bibfnamefont {B.}~\bibnamefont {Favier}}, \bibinfo
  {author} {\bibfnamefont {J.~R.}\ \bibnamefont {Taylor}}, \bibinfo {author}
  {\bibfnamefont {P.~R.}\ \bibnamefont {Holland}}, \ and\ \bibinfo {author}
  {\bibfnamefont {A.}~\bibnamefont {Jenkins}},\ }\href {\doibase
  10.1017/jfm.2020.1064} {\bibfield  {journal} {\bibinfo  {journal} {Journal of
  Fluid Mechanics}\ }\textbf {\bibinfo {volume} {911}},\ \bibinfo {pages} {A44}
  (\bibinfo {year} {2021})}\BibitemShut {NoStop}%
\bibitem [{\citenamefont {Hester}\ \emph {et~al.}(2021)\citenamefont {Hester},
  \citenamefont {McConnochie}, \citenamefont {Cenedese}, \citenamefont
  {Couston},\ and\ \citenamefont {Vasil}}]{Hester:2021}%
  \BibitemOpen
  \bibfield  {author} {\bibinfo {author} {\bibfnamefont {E.~W.}\ \bibnamefont
  {Hester}}, \bibinfo {author} {\bibfnamefont {C.~D.}\ \bibnamefont
  {McConnochie}}, \bibinfo {author} {\bibfnamefont {C.}~\bibnamefont
  {Cenedese}}, \bibinfo {author} {\bibfnamefont {L.-A.}\ \bibnamefont
  {Couston}}, \ and\ \bibinfo {author} {\bibfnamefont {G.}~\bibnamefont
  {Vasil}},\ }\href {\doibase 10.1103/PhysRevFluids.6.023802} {\bibfield
  {journal} {\bibinfo  {journal} {Phys. Rev. Fluids}\ }\textbf {\bibinfo
  {volume} {6}},\ \bibinfo {pages} {023802} (\bibinfo {year}
  {2021})}\BibitemShut {NoStop}%
\bibitem [{\citenamefont {Angot}\ \emph {et~al.}(1999)\citenamefont {Angot},
  \citenamefont {Bruneau},\ and\ \citenamefont {Fabrie}}]{Angot:1999}%
  \BibitemOpen
  \bibfield  {author} {\bibinfo {author} {\bibfnamefont {P.}~\bibnamefont
  {Angot}}, \bibinfo {author} {\bibfnamefont {C.-H.}\ \bibnamefont {Bruneau}},
  \ and\ \bibinfo {author} {\bibfnamefont {P.}~\bibnamefont {Fabrie}},\ }\href
  {\doibase 10.1007/s002110050401} {\bibfield  {journal} {\bibinfo  {journal}
  {Numerische Mathematik}\ }\textbf {\bibinfo {volume} {81}},\ \bibinfo {pages}
  {497} (\bibinfo {year} {1999})}\BibitemShut {NoStop}%
\bibitem [{\citenamefont {Brown}\ \emph {et~al.}(2001)\citenamefont {Brown},
  \citenamefont {Cortez},\ and\ \citenamefont {Minion}}]{Brown:2001}%
  \BibitemOpen
  \bibfield  {author} {\bibinfo {author} {\bibfnamefont {D.~L.}\ \bibnamefont
  {Brown}}, \bibinfo {author} {\bibfnamefont {R.}~\bibnamefont {Cortez}}, \
  and\ \bibinfo {author} {\bibfnamefont {M.~L.}\ \bibnamefont {Minion}},\
  }\href {\doibase https://doi.org/10.1006/jcph.2001.6715} {\bibfield
  {journal} {\bibinfo  {journal} {Journal of Computational Physics}\ }\textbf
  {\bibinfo {volume} {168}},\ \bibinfo {pages} {464} (\bibinfo {year}
  {2001})}\BibitemShut {NoStop}%
\bibitem [{\citenamefont {Huang}\ and\ \citenamefont
  {Moore}(2021)}]{Huang:2021}%
  \BibitemOpen
  \bibfield  {author} {\bibinfo {author} {\bibfnamefont {J.~M.}\ \bibnamefont
  {Huang}}\ and\ \bibinfo {author} {\bibfnamefont {N.~J.}\ \bibnamefont
  {Moore}},\ }\href@noop {} {\enquote {\bibinfo {title} {Morphological
  attractors in natural convective dissolution},}\ } (\bibinfo {year} {2021}),\
  \Eprint {http://arxiv.org/abs/2109.02212} {arXiv:2109.02212} \BibitemShut
  {NoStop}%
\bibitem [{\citenamefont {Chandrasekhar}(1961)}]{Chandrasekhar:1961}%
  \BibitemOpen
  \bibfield  {author} {\bibinfo {author} {\bibfnamefont {S.}~\bibnamefont
  {Chandrasekhar}},\ }\href@noop {} {\emph {\bibinfo {title} {Hydrodynamic and
  Hydromagnetic Stability}}}\ (\bibinfo  {publisher} {Dover},\ \bibinfo {year}
  {1961})\BibitemShut {NoStop}%
\bibitem [{\citenamefont {Drazin}\ and\ \citenamefont
  {Reid}(2004)}]{Drazin:2004}%
  \BibitemOpen
  \bibfield  {author} {\bibinfo {author} {\bibfnamefont {P.~G.}\ \bibnamefont
  {Drazin}}\ and\ \bibinfo {author} {\bibfnamefont {W.~H.}\ \bibnamefont
  {Reid}},\ }\href {\doibase 10.1017/CBO9780511616938} {\emph {\bibinfo {title}
  {Hydrodynamic Stability}}},\ \bibinfo {edition} {2nd}\ ed.,\ Cambridge
  Mathematical Library\ (\bibinfo  {publisher} {Cambridge University Press},\
  \bibinfo {year} {2004})\BibitemShut {NoStop}%
\bibitem [{\citenamefont {Moore}(1978)}]{Moore:1978}%
  \BibitemOpen
  \bibfield  {author} {\bibinfo {author} {\bibfnamefont {D.~W.}\ \bibnamefont
  {Moore}},\ }\href {\doibase https://doi.org/10.1002/sapm1978582119}
  {\bibfield  {journal} {\bibinfo  {journal} {Studies in Applied Mathematics}\
  }\textbf {\bibinfo {volume} {58}},\ \bibinfo {pages} {119} (\bibinfo {year}
  {1978})}\BibitemShut {NoStop}%
\bibitem [{\citenamefont {Dhanak}(1994)}]{Dhanak:1994}%
  \BibitemOpen
  \bibfield  {author} {\bibinfo {author} {\bibfnamefont {M.~R.}\ \bibnamefont
  {Dhanak}},\ }\href {\doibase 10.1017/S0022112094001552} {\bibfield  {journal}
  {\bibinfo  {journal} {Journal of Fluid Mechanics}\ }\textbf {\bibinfo
  {volume} {269}},\ \bibinfo {pages} {265–281} (\bibinfo {year}
  {1994})}\BibitemShut {NoStop}%
\bibitem [{\citenamefont {Grossmann}\ and\ \citenamefont
  {Lohse}(2000)}]{Grossmann:2000}%
  \BibitemOpen
  \bibfield  {author} {\bibinfo {author} {\bibfnamefont {S.}~\bibnamefont
  {Grossmann}}\ and\ \bibinfo {author} {\bibfnamefont {D.}~\bibnamefont
  {Lohse}},\ }\href {\doibase 10.1017/S0022112099007545} {\bibfield  {journal}
  {\bibinfo  {journal} {Journal of Fluid Mechanics}\ }\textbf {\bibinfo
  {volume} {407}},\ \bibinfo {pages} {27–56} (\bibinfo {year}
  {2000})}\BibitemShut {NoStop}%
\bibitem [{\citenamefont {Qiu}\ and\ \citenamefont {Tong}(2001)}]{Qiu:2001}%
  \BibitemOpen
  \bibfield  {author} {\bibinfo {author} {\bibfnamefont {X.-L.}\ \bibnamefont
  {Qiu}}\ and\ \bibinfo {author} {\bibfnamefont {P.}~\bibnamefont {Tong}},\
  }\href {\doibase 10.1103/PhysRevE.64.036304} {\bibfield  {journal} {\bibinfo
  {journal} {Phys. Rev. E}\ }\textbf {\bibinfo {volume} {64}},\ \bibinfo
  {pages} {036304} (\bibinfo {year} {2001})}\BibitemShut {NoStop}%
\bibitem [{\citenamefont {Grossmann}\ and\ \citenamefont
  {Lohse}(2002)}]{Grossmann:2002}%
  \BibitemOpen
  \bibfield  {author} {\bibinfo {author} {\bibfnamefont {S.}~\bibnamefont
  {Grossmann}}\ and\ \bibinfo {author} {\bibfnamefont {D.}~\bibnamefont
  {Lohse}},\ }\href {\doibase 10.1103/PhysRevE.66.016305} {\bibfield  {journal}
  {\bibinfo  {journal} {Phys. Rev. E}\ }\textbf {\bibinfo {volume} {66}},\
  \bibinfo {pages} {016305} (\bibinfo {year} {2002})}\BibitemShut {NoStop}%
\bibitem [{\citenamefont {Moon}\ \emph {et~al.}(2018)\citenamefont {Moon},
  \citenamefont {Sutherland}, \citenamefont {Carroll}, \citenamefont
  {Felikson}, \citenamefont {Kehrl},\ and\ \citenamefont
  {Straneo}}]{Moon:2018}%
  \BibitemOpen
  \bibfield  {author} {\bibinfo {author} {\bibfnamefont {T.}~\bibnamefont
  {Moon}}, \bibinfo {author} {\bibfnamefont {D.~A.}\ \bibnamefont
  {Sutherland}}, \bibinfo {author} {\bibfnamefont {D.}~\bibnamefont {Carroll}},
  \bibinfo {author} {\bibfnamefont {D.}~\bibnamefont {Felikson}}, \bibinfo
  {author} {\bibfnamefont {L.}~\bibnamefont {Kehrl}}, \ and\ \bibinfo {author}
  {\bibfnamefont {F.}~\bibnamefont {Straneo}},\ }\href {\doibase
  10.1038/s41561-017-0018-z} {\bibfield  {journal} {\bibinfo  {journal} {Nature
  Geoscience}\ }\textbf {\bibinfo {volume} {11}} (\bibinfo {year} {2018}),\
  10.1038/s41561-017-0018-z}\BibitemShut {NoStop}%
\bibitem [{\citenamefont {Brostr{\"o}m}\ \emph {et~al.}(2009)\citenamefont
  {Brostr{\"o}m}, \citenamefont {Melsom}, \citenamefont {Sayed},\ and\
  \citenamefont {Kubat}}]{brostrom2009iceberg}%
  \BibitemOpen
  \bibfield  {author} {\bibinfo {author} {\bibfnamefont {G.}~\bibnamefont
  {Brostr{\"o}m}}, \bibinfo {author} {\bibfnamefont {A.}~\bibnamefont
  {Melsom}}, \bibinfo {author} {\bibfnamefont {M.}~\bibnamefont {Sayed}}, \
  and\ \bibinfo {author} {\bibfnamefont {I.}~\bibnamefont {Kubat}},\
  }\href@noop {} {\bibfield  {journal} {\bibinfo  {journal} {Journal of
  Geophysical Research}\ }\textbf {\bibinfo {volume} {99}},\ \bibinfo {pages}
  {3337} (\bibinfo {year} {2009})}\BibitemShut {NoStop}%
\bibitem [{\citenamefont {Claudin}\ \emph {et~al.}(2017)\citenamefont
  {Claudin}, \citenamefont {Durán},\ and\ \citenamefont
  {Andreotti}}]{Claudin:2017}%
  \BibitemOpen
  \bibfield  {author} {\bibinfo {author} {\bibfnamefont {P.}~\bibnamefont
  {Claudin}}, \bibinfo {author} {\bibfnamefont {O.}~\bibnamefont {Durán}}, \
  and\ \bibinfo {author} {\bibfnamefont {B.}~\bibnamefont {Andreotti}},\ }\href
  {\doibase 10.1017/jfm.2017.711} {\bibfield  {journal} {\bibinfo  {journal}
  {Journal of Fluid Mechanics}\ }\textbf {\bibinfo {volume} {832}},\ \bibinfo
  {pages} {R2} (\bibinfo {year} {2017})}\BibitemShut {NoStop}%
\bibitem [{\citenamefont {Bushuk}\ \emph {et~al.}(2019)\citenamefont {Bushuk},
  \citenamefont {Holland}, \citenamefont {Stanton}, \citenamefont {Stern},\
  and\ \citenamefont {Gray}}]{Bushuk:2019}%
  \BibitemOpen
  \bibfield  {author} {\bibinfo {author} {\bibfnamefont {M.}~\bibnamefont
  {Bushuk}}, \bibinfo {author} {\bibfnamefont {D.~M.}\ \bibnamefont {Holland}},
  \bibinfo {author} {\bibfnamefont {T.~P.}\ \bibnamefont {Stanton}}, \bibinfo
  {author} {\bibfnamefont {A.}~\bibnamefont {Stern}}, \ and\ \bibinfo {author}
  {\bibfnamefont {C.}~\bibnamefont {Gray}},\ }\href {\doibase
  10.1017/jfm.2019.398} {\bibfield  {journal} {\bibinfo  {journal} {Journal of
  Fluid Mechanics}\ }\textbf {\bibinfo {volume} {873}},\ \bibinfo {pages}
  {942–976} (\bibinfo {year} {2019})}\BibitemShut {NoStop}%
\bibitem [{\citenamefont {Josberger}\ and\ \citenamefont
  {Martin}(1981)}]{Josberger:1981}%
  \BibitemOpen
  \bibfield  {author} {\bibinfo {author} {\bibfnamefont {E.~G.}\ \bibnamefont
  {Josberger}}\ and\ \bibinfo {author} {\bibfnamefont {S.}~\bibnamefont
  {Martin}},\ }\href {\doibase 10.1017/S0022112081002450} {\bibfield  {journal}
  {\bibinfo  {journal} {Journal of Fluid Mechanics}\ }\textbf {\bibinfo
  {volume} {111}},\ \bibinfo {pages} {439–473} (\bibinfo {year}
  {1981})}\BibitemShut {NoStop}%
\end{thebibliography}%
\bibliographystyle{apsrev4-1}

\end{document}